\magnification=1200
\def\nonumfirst{\nopagenumbers
                \footline={\ifnum\count0=1\hfill
                           \else\hfill\folio\hfill
                           \fi}}
\nonumfirst
\magnification=1200
\def\singlespace{\baselineskip 12 pt}

\def\oneandahalfspace{\baselineskip 18pt}
\def\blankline{\vskip 12 pt\noindent}

\def\secto#1\endsecto{\vskip 20pt {\bf #1}\vskip 7pt\nobreak}
\global\newcount\refno \global\refno=1
\newwrite\rfile
\def\ref#1#2{\hbox{[\hskip 2pt\the\refno\hskip 2pt]}\nref#1{#2}}
\def\nref#1#2{\xdef#1{\hbox{[\hskip 2pt\the\refno\hskip 2pt]}}%
\ifnum\refno=1\immediate\openout\rfile=refs.tmp\fi%
\immediate\write\rfile{\noexpand\item{\noexpand#1\ }#2.}%
\global\advance\refno by1}
\def\semi{;\hfil\noexpand\break}
\def\demi{:\hfil\noexpand\break}

\newwrite\efile \let\firsteqn=T
\def\writeqno#1%
{\if T\firsteqn \immediate\openout\efile=eqns.tmp\global\let\firsteqn=F\fi%
\immediate\write\efile{#1 \string#1}\global\advance\meqno by1}

\def\eqnn#1{\xdef #1{(\the\secno.\the\meqno)}\writeqno#1}
\def\eqna#1{\xdef #1##1{(\the\secno.\the\meqno##1)}\writeqno{#1{}}}

\def\eqn#1#2{\xdef #1{(\the\secno.\the\meqno)}%
$$#2\eqno(\the\secno.\the\meqno)$$\writeqno#1}
\def\nobreak{\penalty1000}
\def\titl#1\endtitl{\par\vfil
     \vbox to 2in {}{\bf #1}\nobreak}
\def\titol#1\endtitol{\par\vfil
     \par\vbox to 1in {}{\bf #1}\par\vskip 1in\nobreak}
\def\tit#1\endtit{
     \vbox to 0.5in {}{\bf #1}\nobreak}

%
%
%
%
%
%
%
\def\lspace{\ifx\answ\bigans{}\else\qquad\fi}
\def\lbspace{\ifx\answ\bigans{}\else\hskip-.2in\fi} 
%
%
%

\def\CM{{\cal M}}

%
%
%
%

%

\def\ket#1{\left| #1\right\rangle}

%
%
\def\frac#1#2{{\textstyle{#1\over #2}}} 
%
%
%
%
\def\tr{\mathop{\rm tr}}

\def\Im{\mathop{\rm Im}}
\def\Re{\mathop{\rm Re}}

\def\kok{ {\rm K}^ 0 - \overline{\rm K}{}^0 }
\def\bob{ {\rm B}^ 0 - \overline{\rm B}{}^0 }
\def\dod{ {\rm D}^ 0 - \overline{\rm D}{}^0 }

\def\ko{ {\rm K}^ 0 }
\def\kob{ \overline{\rm K}{}^0 }
\def\kl{ {\rm K_{L}} }
\def\ks{ {\rm K_{S}} }
\def\ku{ {\rm K_{1}} }
\def\kd{ {\rm K_{2}} }

\def\gl{ {\it \Gamma}_{\rm L} }
\def\kos{ | {\rm K^ 0} \rangle }
\def\kobs{ | {\rm \overline{K}{}^0} \rangle }
\def\bo{ {\rm B}^ 0 }
\def\bbo{ \overline{\rm B}{}^0 }
\def\BB{\hbox{B\=B}}

\def\do{ {\rm D}^ 0 }
%
%
%
\def\ltap{\ \raise.3ex\hbox{$<$\kern-.75em\lower1ex\hbox{$\sim$}}\ }
\def\gtap{\ \raise.3ex\hbox{$>$\kern-.75em\lower1ex\hbox{$\sim$}}\ }
\def\gl{\ \raise.5ex\hbox{$>$}\kern-.8em\lower.5ex\hbox{$<$}\ }
\def\roughly#1{\raise.3ex\hbox{$#1$\kern-.75em\lower1ex\hbox{$\sim$}}}
%
%

%
\def\[{\left[}
\def\]{\right]}
\def\({\left(}
\def\){\right)}
%
%

\def\nc{{Nuovo Cimento}\ }

\def\pl{{Phys. Lett.}\ }
\def\pr{{Phys. Rev.}\ }

\def\prl{{Phys. Rev. Lett.}\ }

%
\textfont2=\tensy \scriptfont2=\sevensy \scriptscriptfont2=\fivesy
\def\cal{\fam2}
\def\Iscr{{\cal I}}

\def\Qscr{{\cal Q}}
\def\Rscr{{\cal R}}
\def\Hscr{{\cal H}}
\def\Gscr{{\cal G}}
\def\Cscr{{\cal C}}

\def\Uscr{{\cal U}}
\def\Vscr{{\cal V}}
\def\Sscr{{\cal S}}
%

\def\st{\scriptstyle}

\def\pmb#1{\setbox0=\hbox{$#1$}%
  \kern-.025em\copy0\kern-\wd0
  \kern.05em\copy0\kern-\wd0
  \kern-.025em\raise.0433em\box0}
\def\pmbs#1{\setbox0=\hbox{$\st #1$}%
  \kern-.0175em\copy0\kern-\wd0
  \kern.035em\copy0\kern-\wd0
  \kern-.0175em\raise.0303em\box0}

\def\bfs#1{\hbox to .0035in{$\st#1$\hss}\hbox to .0035in{$\st#1$\hss}\st#1}

\def\bfDelta{\pmb{\Delta}}
\def\bfPi{\pmb{\Pi}}
\def\bfLambda{\pmb{\Lambda}}
\def\bfDeltap{\pmb{\Delta^\prime}}

\def\bfI{\hbox{\bf I}}

\def\bfR{\hbox{\bf R}}

%

\def\ket#1{\vert #1 \rangle }

%
%
\global\newcount\meqno \global\meqno=1
\newwrite\efile \let\firsteqn=T
\def\writeqno#1%
{\if T\firsteqn \immediate\openout\efile=eqns.tmp\global\let\firsteqn=F\fi%
\immediate\write\efile{#1 \string#1}\global\advance\meqno by1}

\def\eqqn#1#2{\xdef #1{(\the\meqno)}%
$$#2\eqno(\the\meqno)$$\writeqno#1}
\font\medf=cmb10 scaled \magstep3
%
\vsize=25 truecm
\hsize=16 truecm
\voffset=-0.8 truecm
%
\singlespace
\parskip 6truept
\parindent 20truept
\vbox{ {\rightline{\bf IFUM 556-FT/97}}
       {\rightline{\bf UNIBAS--TH 7/97}}
}
\hyphenation{ex-pe-ri-men-tal}
\hyphenation{va-cu-um}
\vskip 5truecm
\centerline{\medf 
		The Scattering Theory of CP Violation}
\vskip 2truecm
\vskip 33truept
\centerline{Decio Cocolicchio}
\vskip 20truept
\vbox{
\centerline{\it Dipartimento di Matematica,
Univ. Basilicata, Potenza, Italy}
\vskip 5truept
\centerline{\it Via N. Sauro 85, 85100 Potenza, Italy} }
\vskip 15truept
\vbox{
\centerline{\it Istituto Nazionale di Fisica Nucleare,
                     Sezione di Milano, Italy}
\vskip 5truept
\centerline{\it Via G. Celoria 16, 20133 Milano, Italy} }
\vskip 4truecm
\centerline{\it ABSTRACT}
\vskip 15truept
\singlespace
\noindent
The mixing effects and the $CP(T)$--violating formalism for Kaon,
$B$--mesons and similar unstable oscillating systems,
are recovered by means of a method based on the properties of the complex
singularities in the $S$--matrix theory with unstable intermediate states.
General $q^2$--dependent relations for $CP$--asymmetries are then introduced.
\vskip 1truecm \noindent \singlespace
\vbox{
      {\leftline{IFUM 556-FT/97}}
      {\leftline{UNIBAS--TH 7/97}}
      }
\vfill\eject
\vsize=24 truecm
\hsize=16 truecm
\baselineskip 18 truept
\parindent=1cm
\parskip=8pt
\oneandahalfspace
%
%
\phantom{.}
\blankline
\leftline {\bf I. Introduction}
\blankline
\noindent
Recently, there has been a renewed interest in the theoretical approach 
to neutral meson systems like $\kok$, $\bob$ or $\dod$, due 
mainly to the planned advances in the experimental methods
in DA$\Phi$NE project at Frascati (Rome) and BABAR at SLAC (Stanford)
to measure their small observables. Therefore, it becomes important to 
predict carefully these tiny effects. 
From the most general point of 
view of non relativistic Quantum Mechanics, the problem consists in 
defining correctly the position dependent probability which let us 
distinguish the factorized constituents ($\ko$ or $\kob$ for example,
but they can be also $\bo$ or $\do$ systems)
of the entangled mass eigenstates ${\rm K_{L,S}}$ or $CP$-eigenstates
${\rm K_{1,2}}$.
In doing so, usually one restricts to a single pole (Lee-Oehme-Yang 
or LOY) approximation
\ref\LOY{T. D. Lee, R. Oehme and C. N. Yang, Phys. Rev. {\bf 106} 
(1957) 340; P. K. Kabir, {\it The CP Puzzle} (Academic Press,
New York, 1968), Appendix A} \ 
to describe the temporal evolution of the neutral kaon complex.
The controversial point of the validity of this approximation consists 
in the fact that the characteristic time dependence of the survival 
probability of any metastable state $\ket{\phi}$
\eqqn\eIi{
P(t) = \vert \alpha (t) \vert^2 = \vert \langle \phi \vert 
e^{-iHt} \vert \phi \rangle \vert^2 = \int e^{-iEt} \rho(E) dE
}
\noindent
can deviate from a pure exponential in the case that the decay energy 
spectrum
\eqqn\eIii{
\rho(E) =\vert\langle\psi_E\vert\phi\rangle\vert^2
}
\noindent
is unbounded, being $\ket{\psi_E}$ the eigenstate of an effective 
Hamiltonian: $H\ket{\psi_E} = E \ket{\psi_E}$.
In recent years, such a problem has been discussed extensively in the 
case of the neutral kaon system
\ref\KAK{
L. A. Khalfin, ``{\it New Results on the $CP$--violation Problem}'',
Univ. Texas Report DOE-ER40200-211 (February 1990);
C.B. Chiu and E.C.G. Sudarshan, Phys. Rev. {\bf D42} (1990) 3712;
Ya. I. Azimov, JETP Lett. {\bf 58} (1993) 159;
K. Urbanowsky, Int. J. Mod. Phys. {\bf A10} (1995) 1151},
with controversial results about the eventual physical effects of the
regeneration of the physical kaons in the vacuum as due to the off-diagonal 
quenching of the resonant contributions.
Nevertheless, since we are dealing with an approximate theory, it is 
not surprising that one expects departures from its predictions and an 
update of the LOY formalism seems needed
\ref\BMT{C. Bernardini, L. Maiani and L. Testa, \prl {\bf 71} (1993) 
2687}. Deviations from the exponential decay law in the time evolution 
of any metastable system is expected at very short or very long times 
as compared to the lifetime of the unstable particle 
\ref\Instab{D. Cocolicchio, 
``{\it The Characterization of Unstable Particles}'',
preprint IFUM 514/FT-96, Istituto Nazionale di Fisica Nucleare,
Sezione di Milano, submitted to Physical Review D } \ 
and in dependence of the structure of the prepared initial state 
\ref\JS{J. Schwinger, Ann. Phys. {\bf 9} (1960) 169}.
These basic features can be introduced in the formalism of the 
density matrix with the generalization of the Liouville master 
equation which governs the decay and the evolution of the neutral kaon 
system and, in the same time, can distinguish the factorized 
constituents $K_{L,S}$ of the entangled state $\kok$
\ref\QMCPT{
J. Ellis, J.S. Hagelin, D.V. Nanopoulos and M. Srednicki,
Nucl Phys. {\bf B241} (1984) 381;
P. Huet and M.E. Peskin, Nucl. Phys. {\bf B434} (1995) 3;
J. Ellis, J.L. Lopez, N.E. Mavromatos and D.V. Nanopoulos, 
Phys. Rev. {\bf D53} (1996) 3846;
F. Benatti and R. Floreanini, Phys. Lett. {\bf B389} (1996) 100}.
An alternative approach expresses the survival amplitude in terms of 
the relative propagator function and consider the non relativistic 
limit of the space-time evolution in terms of the coherent overlapping 
of two Wightmann expanded wave-packets which somehow mimics the 
interference effects of two kaons
\ref\SWS{E. Sassaroli, Y.N. Srivastava and A. Widom, \pl {\bf 344B} 
(1995) 436; H. J. Lipkin, Phys. Lett. {\bf B348} (1995) 604}.
This treatment advocated the presence of two different energies and 
momenta in order to account for the separation between the $\ks$ and 
the $\kl$ states. 
Moreover, the breakdown of the exponential decay law can be noted 
also in the context of $S$-matrix scattering field theory
\ref\SVelt{
P.T. Matthews and A. Salam, 
\pr {\bf 115} (1959) 1079;  
M. Levy, \nc {\bf 13} (1959) 115; {\bf 14} (1959) 612;
R. Jacob and R.G. Sachs, \pr {\bf 121} (1961) 350;    
M. Veltman, Physica {\bf 29} (1963) 186;
H. P. Stapp, Nuovo Cimento, {\bf 32} (1964) 103; 
J. Gunson, J. Math. Phys. {\bf 6} (1965) 827, 845, 852;
R. Eden, P. Landshoff, D. Olive and J. Polkinghorne, ``{\it The Analytic 
$S$-matrix}'', (Cambridge Univ. Press, Cambridge, 1966)}.
In this context, the correct treatment of unstable oscillating systems,
and our present understanding of the $CP$--violating effects, 
seems to be described by several formalisms:
including the Wigner $R$--matrix approach (mainly adopted in nuclear theory
\ref\Rmat{A. M. Lane and R. G. Thomas, Rev. Mod. Phys. {\bf 30} (1958) 
257; Y. Yamaguchi, J. Phys. Soc. Japan {\bf 58} (1989) 4375; {\bf 60}
(1991) 1541;
V.V. Sokolov and V. G. Zelevinsky, Nucl. Phys. {\bf A504} (1989) 562}), 
the $K$-matrix technique
\ref\Kmat{
R. L. Warnock, Ann. Phys. (NY) {\bf 65} (1971) 386; 
R. G. Newton, ``{\it Scattering Theory of Waves and Particles}'',
(McGraw-Hill, New York, 1966);
A.N. Kamal, N. R. Sinha, Z. Phys. {\bf C41} (1988) 207;
M. Wanninger, L.M. Sehgal, Z. Phys. {\bf C50} (1991) 47},
and methods based purely on the framework of the $S$-matrix perturbation theory
\ref\Smat{
G. C. Wick, Phys. Lett. {\bf 30B} (1969) 126;
L. Wolfenstein, Phys. Rev. {\bf 188} (1969) 2536;
Y. Dothan and D. Horn, Phys. Rev. {\bf D1} (1970) 916;
L. Stodolsky, Phys. Rev. {\bf D1} (1970) 2683}.
The correct quantum field treatment of unstable particles
within the framework of the $S$-matrix perturbation 
theory~\Instab , is motivated mainly by the strength to generalize the 
Breit-Wigner propagator with the inclusion of the resummation of higher order 
quantum corrections, which often results in several pathologies in 
gauge field theories. Even though this approach may eventually 
furnishes gauge invariant results, nevertheless, the perturbation treatment 
introduces residual threshold terms which become more effective when 
$CP$--violating effects are considered
\ref\BWW{A. Pilaftsis, Z. Phys. {\bf C47} (1990) 95;
J. Liu and G. Segre, Phys. Rev. {\bf D49} (1994) 1342;
J. Papavassiliou and A. Pilaftsis, Phys. Rev. {\bf D53} (1996) 2128}.
The advanced features of today experimental methods suggest then to improve 
the phenomenological assumptions for those 
observables for which the $CP$--violating 
contributions are significant. One of the most
sensible physical ground is the entangled $\kok$ interference amplitude.
Although, its observables rest inextricably dominated by mixing 
(i.e. negligible ``direct'' $K\rightarrow (\pi\pi)_I$ contributions
\ref\SM{
A. J. Buras, ``{\it CP Violation: Present and 
Future}'', Proceedings 1st Int. Conf. on Phenomenology of Unification, 
Rome, 1994} ),
the assumption that, in general, the mixing of elementary particles
is independent of the momentum squared of the underlying intermediate 
channels is more problematic, mainly in the case of the vector modes
\ref\qdepen{S. Coleman and H. Schnitzer, Phys. Rev. {\bf 134} (1964) B863;
R. G. Sachs and J. F. Willemsen, Phys. Rev. {\bf D2} (1970) 133;
F. M. Renard, Springer Tracts Mod. Phys. {\bf 63} (1972) 98}.
Width effects, for instance, have been suggested as a possible 
mechanism for generating resonant $CP$--violating contributions that 
could provide a window into new physics
\ref\RCPV{
R. Cruz, B. Grzakowski and J. F. Gunion, Phys. Lett. {\bf B289} (1992) 440; 
D. Atwood, G. Eilam, A. Soni, R. R. Mendel and R. Migneron, 
Phys. Rev. Lett. {\bf 70} (1993) 1364;
J. Liu, Phys. Rev. {\bf D47} (1993) R1741;
M. Nowakowski and A. Pilaftsis, Z. Phys. {\bf C60} (1993) 121;
U. Baur and D. Zeppenfeld, Phys. Rev. lett. {\bf 75} (1995) 1002;
T. Arens and L. M. Sehgal, Phys. Rev. {\bf D51} (1995) 3525}.
However, the inclusion of the precise $q^2$--dependence, usually 
thought to be small, could have unexpectedly large effects on the 
extraction of those sensible observables which give significant 
contributions to $CP$-asymmetries.
Nevertheless, the usual effective complex mass matrix methods of the narrow 
width LOY approximation neglects this effect, and also it rests 
completely unappropriate to implement 
the notion of the rest frame for an 
oscillating unstable composite system.
All that, in turn, seems to require an explicit relativistic 
description of the kaon system, also in order to avoid the phase 
ambiguities that arise when the superposition of states with different 
mass and momentum are described in a Galilean invariant form
\ref\Galinv{V. Bargmann, Ann. Phys. {\bf 54} (1954) 1;
A. W. Wightmann, Rev. Mod. Phys. {\bf 34} (1962) 845;
J. M. Levy-Leblond, J. Math. Phys. {\bf 4} (1963) 776 and Comm. Math. 
Phys. {\bf 4} (1967) 157;
E. Henley and W. Thirring, {\it Elementary Quantum Field
Theory}, McGraw Hill, New York (1963)}.
Indeed, the essential content of the
time dependent properties of the kaon complex can be introduced
without recourse to the S-matrix formalism, by considering simply the
subtleties related to the
location of the complex singularities 
in the multiple sheets of the Riemann 
surface into which the Fourier transform of the propagator can be 
continued analytically. Such propagator's method has the great 
advantage to appear natural and indeed independent of various 
production and decay mechanisms, although it is not immediate to
have a model and to solve the ambiguities 
connected to its complex analytical structure.
The propagator formalism for the $\kok$ system
\ref\Kpropag{R. G. Sachs, Ann. Phys. {\bf 22} (1963) 239;
J. Harte and R. G. Sachs, Phys. Rev. {\bf 135} (1964) B459;
R. G. Sachs, {\it The Physics of Time Reversal
Invariance} (University of Chicago Press, Chicago, 1988);
O. Nachtmann, Acta Phys. Austr. Suppl. {\bf 6} (1969) 485;
D. Sudarsky, E. Fishbach, C. Talmadge, S. H. Aronson and H. Y. Cheng,
Ann. Phys. {\bf 207} (1991) 103} \ 
results a rigorous relativistic treatment which arises naturally in 
the context of quantum field theory.
In this paper, we propose to derive some general properties of kaon 
mixing in some detail, focusing on superweak CP-violating processes and 
combining the dynamics of the complex pole of the kaon field 
propagator with the results of the spectral formalism of the time 
dependent perturbation method.
\blankline
\leftline {\bf II. 
The Kaon Complex Within and Beyond the Wigner--Weisskopf}
\leftline {\phantom{\bf II. }\bf Narrow Width Approximation.}

\noindent
In this section, we investigate the shortcomings of the 
Lee-Oehme-Yang theory~\LOY \ of the decay and evolution of the neutral 
kaon system.
It is characteristic of the Wigner--Weisskopf narrow width 
approximation that the effective Hamiltonian acts upon a Hilbert 
subspace in which the exponential decay law is assured in order to 
generate a dynamical semigroup evolution.
The time evolution of the flavour states
\blankline
\eqqn\eIIu{
\left( \matrix{ \vert \ko (t)  \rangle \cr 
                    \cr
                \vert \kob (t) \rangle \cr}\right) 
=  \Uscr (t) 
\left( \matrix{ \vert \ko \rangle \cr 
                 \cr
          \vert \kob \rangle \cr} \right) \quad ,
}
and similarly for the mass right--eigenstates
\blankline
\eqqn\eIId{
\left( \matrix{ \vert K_S (t) \rangle \cr 
                 \cr
                \vert K_L (t) \rangle \cr}\right) 
=  \Vscr (t) 
\left( \matrix{ \vert K_S \rangle \cr 
                 \cr
                \vert K_L \rangle \cr}\right) \quad 
}
are governed by the matrix elements
\eqqn\eIIt{
\eqalign{
U_{ij} =& \langle K_i \vert 
\exp \left[ -{i\over {\hbar} } \Hscr t\right] 
\vert K_j \rangle \cr
& \cr
V_{\alpha\beta} =& \langle K_\alpha^\prime \vert 
\exp \left[ -{i\over {\hbar} } \Hscr t\right] 
\vert K_\beta \rangle \cr}
}
where the latin indices denote
$\ko$, $\kob$ and the greek letters denote $K_S$, $K_L$, respectively.
The evolution matrices $\Uscr$ 
and $\Vscr$ are then related by the following similarity transformation
\eqqn\eIIq{
\Uscr=\Rscr \Vscr \Rscr^{-1} 
}
The effective Hamiltonian matrix is determined by eight real 
parameters, but only seven are physical meaningful because the 
absolute phase of $H_{12}$ or $H_{21}$ is meaningless, being the 
relative phase of $\kos$ and $\kobs$ arbitrary.
They can be substituted by the two complex eigenvalues
\eqqn\eIIad{
\lambda_S = {1\over 2} \left( C - D \right) }
\eqqn\eIIadb{
\lambda_L = {1\over 2} \left( C + D \right) }
where, in a general theory,
$C =  \lambda_L +\lambda_S = H_{11} + H_{22} =\tr \Hscr $ and
$D^2 = (\lambda_L -\lambda_S)^2 = \left(H_{11} - H_{22}\right)^2 
+ 4 H_{12} H_{21} = (\tr \Hscr)^2 - 4 (\det \Hscr)$,
and the two complex mixing parameters $\epsilon_{S, L}$
are given by:
\eqqn\eIIepsl{
\eqalign{
\epsilon_S =& 
\left({ {2 H_{12} - D}
             \over
        {2 H_{12} + D} }\right)  -
\left({ {4 H_{12}}
              \over
         {2 H_{12} + D} }\right) 
\left({  {H_{11} - H_{22}}
             \over
         {H_{11} - H_{22} + D + 2 H_{12} } }\right) = \epsilon - \delta_S \cr
\epsilon_L =& 
\left({ {2 H_{12} - D}
             \over
         {2 H_{12} + D} }\right)  -
\left({ {4 H_{12}}
              \over
         {2 H_{12} + D} }\right) 
\left({  {H_{11} - H_{22}}
             \over
         {H_{11} - H_{22} - D - 2 H_{12}} }\right) = \epsilon - \delta_L \cr}
}
where

\eqqn\eIIeps{
\eqalign{
\epsilon = &  { {2 H_{12} - D}
              \over
               { 2 H_{12} + D}   } 
          = {  { \sqrt{H_{12}} - \sqrt{H_{21}} } 
                           \over
               { \sqrt{H_{12}} + \sqrt{H_{21}} } } 
=  {1 \over {i D}} \left({ - \Im M_{12} + i Im \Gamma_{12} }\right) 
\cr
\delta_S =& 
\left({ { 2 H_{12}}\over {2 H_{12} + D} }\right)
\left({ {H_{11} - H_{22} }
                                                    \over
{H_{11} + H_{12} -\lambda_S} }\right)
\cr
\delta_L =& 
\left({ { 2 H_{12}}\over {2 H_{12} + D} } \right)
\left({ {H_{11} - H_{22} }
                                                    \over
        {H_{11} - H_{12} -\lambda_L} }\right)
\cr} }
The complex scaling matrix $\Rscr$, which
diagonalizes the effective Hamiltonian,
is then given by
\eqqn\eIIau{
\eqalign{
\Rscr = & \left(\matrix{  a_S & a_L \cr
             -{ {\displaystyle a_S}   
                     \over
                {\displaystyle 2 H_{12}} }
({\displaystyle H_{11} - H_{22} + D}) &
             -{ {\displaystyle a_L}   
                     \over
                {\displaystyle 2 H_{12}} }
({\displaystyle H_{11} - H_{22} - D})
\cr} \right) \cr
& \cr
& = \left(\matrix{
N_S (1+\epsilon_S)  & N_L (1+\epsilon_L) \cr
-N_S (1-\epsilon_S) & \phantom{-}N_L(1-\epsilon_L)\cr}\right)
= \left(\matrix{ p & p^\prime \cr
                 - q & q^\prime \cr}\right)
\, ,\qquad \cr}
}
where $a_S$ and $a_L$ are fixed only once the eigenvectors
normalization is realized and
\eqqn\eIIeta{
\eqalign{
\eta_S =&{ {- q}\over p} = - { {(1-\epsilon_S)}
                  \over
             {(1+\epsilon_S)} }
       = - { {(H_{11} - H_{22} + D)}
                    \over
              {2 H_{12}} }
       = \phantom{-}{    {2 H_{21}}
                 \over
           {(H_{11} - H_{22} - D)} } \cr
\eta_L =& { { q^\prime}\over {p^\prime}} =\phantom{-} { {(1-\epsilon_L)}
                  \over
             {(1+\epsilon_L)} }
       = - { {(H_{11} - H_{22} - D)}
                    \over
              {2 H_{12} } }
       = {    {2 H_{21}}
                 \over
           { (H_{11} - H_{22} + D) } } \cr }
}
being $N_{S,L}^{-2}= 2(1+|\epsilon_{S,L} |^2)$.
In any $CPT$--invariant theory,
$H_{11} = H_{22}$ and there is only one mixing parameter 
$\epsilon=\epsilon_S=\epsilon_L$ and therefore $\eta_S = -\eta_L$,
$N_S^{-2}=N_L^{-2}=N^{-2}=2(1+|\epsilon |^2)$, $p=p^\prime$, $q=q^\prime$.
In this case, we have that
\eqqn\eIIaad{
\lambda_S = H_{11} -\sqrt {H_{12}H_{21}}=M_{11}-{i\over 2} \Gamma_{11}
-{D\over 2}=m_S -{i\over 2} \gamma_S
}
\eqqn\eIIaat{
\lambda_L = H_{11} +\sqrt {H_{12}H_{21}}=M_{11}-{i\over 2} \Gamma_{11}
+{D\over 2}=m_L -{i\over 2} \gamma_L 
}
with
\eqqn\eIIaaq{
D = 2 \sqrt{ H_{12} H_{21} } = 2
\sqrt{(M_{12}-{i\over 2} \Gamma_{12}) ({M_{12}}^*-{i\over 2} 
{\Gamma_{12}}^*)}=\left(\Delta m - {i\over 2} \Delta\gamma\right) .
}
These real ($m_{S,L}$) and imaginary ($\gamma_{S,L}$)
components will define the 
masses and the decay widths of the ${\cal H}$ eigenstates $K_S$ and $K_L$
in the narrow width approximation. 
These short- and long-lived particles result then a linear combination 
of the flavour $\ko$ and $\kob$ states:
\eqqn\eIIo{
\left( \matrix{ \vert K_S \rangle \cr 
            \cr
          \vert K_L \rangle \cr}\right) =
\Rscr^t
\left( \matrix{ \vert \ko \rangle \cr 
     \cr     
     \vert \kob \rangle \cr}\right) \; ,
}
where usually $\Rscr^t$ is preferably parameterized
according to the following relations
\eqqn\eIIse{
\Rscr^t =  
 {1\over {\sqrt{2(1+\vert\epsilon\vert^2)}}} 
\left(
\matrix{
(1+\epsilon) & -(1-\epsilon) \cr
(1+\epsilon) &  (1-\epsilon) \cr}
\right) =
{1\over{\sqrt{1+\vert\eta\vert^2}}}
\left( \matrix{ 1 & \eta \cr
         1 & -\eta \cr}
\right)
= \left(
\matrix{
p & -q \cr
p & q \cr}
\right)
\quad.
}
\noindent
After corresponding normalization of the eigenvectors
\eqqn\eIIn{
\langle K_L \vert K_L \rangle = \langle K_S \vert K_S \rangle
= |p|^2 + |q|^2 = 1\quad ,
}
the impurity parameters can be connected by the simple relations
\eqqn\eIIx{
\epsilon = \frac{\displaystyle p-q }{\displaystyle p+q} 
={  { \displaystyle (\sqrt{H_{12}} - \sqrt{H_{21}}) }
                  \over
    {\displaystyle (\sqrt{H_{12}} + \sqrt{H_{21}}) } }
= i { {\displaystyle \Im M_{12} - {i\over 2} \Im \Gamma_{12} }
                  \over
      {\displaystyle \Re M_{12} - {i\over 2} \Re \Gamma_{12} + {D\over 2} } }
\quad .
}
We remember that the phases of $p$, $q$ may be altered by redefining 
the phases of the $K$ states. To the extent that $M_{12}$ and 
$\Gamma_{12}$ share the same phase: $M_{12} =\vert M_{12} \vert 
e^{i\varphi}$, $\Gamma_{12}= \vert \Gamma_{12}\vert e^{i\varphi}$,
there is no $CP$-violation, since $\epsilon= i  
(\sin\varphi)/(1+\cos\varphi)$.
In general, by means of a suitable 
choice of the relative phase between $\ko$ and $\kob$ one can make 
$\Gamma_{12}$ real negative. 
Anyway, both $p$ and $q$ are not 
measurable quantities, whereas the overlap between $\ks$ and $\kl$

\eqqn\eIInn{
\langle K_S \vert K_L \rangle = 
{ {2 \Re \epsilon}
     \over
  {1 + |\epsilon|^2} } = 
 { {1 -\vert\eta\vert^2}
      \over
   {1+\vert\eta\vert^2} } 
\simeq {  {2 z}\over {4 z^2 + 1}}\arg(M_{12})
\quad ,
}
is independent of any phase convention, with
$z={ {\vert M_{12} \vert} \over {\vert \Gamma_{12} \vert} 
}=(0.477\pm 0.003)$.
Nevertheless, the magnitude of
\eqqn\eIInnb{
\eta = -{q\over p} = -{{1-\epsilon}\over{1+\epsilon}} =
-\sqrt{
{ {M_{12}^* - {i\over 2}\Gamma_{12}^*}
      \over
 {M_{12} - {i\over 2}\Gamma_{12}} } }
}
is certainly independent of any phase convention and,
assuming $\Delta\Sscr = \Delta \Qscr$ rule conserved,
it is directly connected to the amount of the kaon semileptonic
charge rate
\eqqn\esl{
A_{SL} =
{{\Gamma(K_L\rightarrow \ell^+\nu X)-\Gamma(K_L\rightarrow \ell^-\nu X)}
\over
{\Gamma(K_L\rightarrow \ell^+\nu X) +\Gamma(K_L\rightarrow \ell^-\nu X)}}
=
{{1-|\eta|^2}\over{1+|\eta|^2}} .
}
Its experimental value is $A_{SL} = (3.27\pm 0.12)\cdot 10^{-3}$
and then the relative phase between $M_{12}$ and $\Gamma_{12}$ results
$(6.53\pm 0.24)\cdot 10^{-3}$.
In the Wu-Yang convention $\hbox{Im}\,\Gamma_{12}=0$, we obtain that
\eqqn\argeps{
\arg(\epsilon)\simeq\cases{
\pi -\Phi_{SW}\quad\hbox{for}\quad\hbox{Im}{M_{12}}>0\cr
\Phi_{SW}\quad\hbox{for}\quad\hbox{Im}{M_{12}} < 0\cr}
}
being the superweak phase $\Phi_{SW}=\tan^{-1}(2 z)$.
Another powerful experimental observable,
which can determine the
$CP$ violation mainly in the $B_d$-meson mixing
\ref\CM{
D. Cocolicchio and L. Maiani, Phys. Lett. {\bf B291} (1992) 155},
in pair decays
\eqqn\epem{
e^+e^-\longrightarrow\Upsilon\longrightarrow\bigl ( \bo\bbo\bigr ) _L
\longrightarrow\bigl ( \ell^-\overline\nu X^+\bigr )
\bigl ( \ell^+\nu Y^-\bigr ) \ .
}
is the dilepton charge asymmetry 
\eqqn\eIInnt{
\eqalign{
A_{CP} = &
{
{ \bigl | \langle\ell^+\ell^+|\Hscr|\BB\rangle\bigr | ^2 -
  \bigl | \langle\ell^+\ell^-|\Hscr|\BB\rangle\bigr | ^2  } 
                      \over
{  \bigl | \langle\ell^+\ell^+|\Hscr|\BB\rangle\bigr | ^2 +
   \bigl | \langle\ell^+\ell^-|\Hscr|\BB\rangle\bigr | ^2 } 
}
= { {\vert\eta\vert^4 - 1} \over {\vert\eta\vert^4 + 1} } \cr
=& - {   {4 \Re\epsilon(1+|\epsilon  |^2)}
            \over
       {(1+|\epsilon |^2)^2 + 4 (\Re\epsilon)^2} } =
\left( {\Im { {\displaystyle \Gamma_{12}}
                     \over
              {\displaystyle M_{12}} 
            }
       }\right)
{1 
\over 
  { 1 + {\displaystyle 1\over 4}
                { {\displaystyle \vert \Gamma_{12}\vert}
                               \over
                  {\displaystyle \vert M_{12}\vert} 
                 } 
   }
}\cr}
}
However, in the $B$ sector, the situation is still unclear, being
$z_B={ {\vert M^B_{12} \vert} \over {\vert \Gamma^B_{12} \vert} }$
very large, $\vert \eta\vert\simeq 1$ and $A_{CP}$ under investigation
\ref\AC{A. Acuto and D. Cocolicchio, Phys. Rev. {\bf D47} (1993) 3945}.
From the known experimental inputs of
$x_d\simeq \Delta m_B\tau_B=(0.67 \pm 0.10)$ and the fact that 
$\Delta m_B\simeq 2  \vert M^B_{12}\vert = (4.0 \pm 0.8)\cdot 10^{-13}$ 
we find that 
$\arg (M^B_{12}) \simeq \left( {{\Im M^B_{12}}\over {\vert 
M_{12}\vert}}\right)$ is in a wide range between 0.03 and 0.73.
It can be easily shown that $\vert \eta\vert =1$ is a 
necessary and sufficient condition for the $CP$ conservation in the 
mixing.
The method, we outlined, appears more transparent from a 
phenomenological point of view, just because its components are 
expressed directly in terms of observables.
Coming back to the specimen situation of the kaon complex,
in order to find the structure of the time evolution matrix elements
$U_{ij}$, we need to generalize the relevant analytic properties 
of the usual quantum theory of 
the reduced resolvent of a linear operator to the case of a non 
Hermitian Hamiltonian. 
Moreover, here, we prefer do not consider a specific dynamical 
model~\KAK\ and we
suppose only to decompose the original non Hermitian Hamiltonian 
into two non commutating part
\eqqn\eIIxiv{
\Hscr = \Hscr_0 + \Hscr_I
}
where $\Hscr_0$ contains the strong interactions under which the kaons 
appear as stable particles and where $\Hscr_I$ is supposed to
induce their decay modes into the 
continuum spectrum of $\Hscr_0$.
If the interaction is small,
the time evolution matrix elements can then be derived
extracting the poles $\lambda_\alpha$ on the second Riemann sheet and 
studying the expansion criteria for these poles dominance in the
spectral decomposition
\eqqn\eIIxi{
U_{ij} =\int_{\hbox{Spec(H)}} d\lambda e^{-i\lambda t} \rho_{ij} 
(\lambda)
}
where the integration extends over the whole spectrum of the 
Hamiltonian and $\rho_{ij} (\lambda)$ is given by
\eqqn\eIIxii{
\rho_{ij} = \langle K_i\vert {\Gscr} \vert K_j \rangle
}
where ${\Gscr} (z) = [ z {\Iscr} - \Hscr ]^{-1} $ rigorously must represent 
the complete reduced resolvent.
Beyond the pole dominance, the details of the result depend of course 
upon the spectral matrix $\rho_{i j}(\lambda)$.
Nevertheless, a sensible spectrum is expected to be bounded
from below and it is possible to normalize the ground state to have 
the zero energy eigenvalue. The integration range of the spectrum then 
extends from 0 to $\infty$.
Despite this cut-off, it is evident that the complete time evolution
of the system is determined in principle by the positivity requirement
of the spectral content of the initial state.
The Wigner Weisskopf approximation amounts to assume that the spectrum 
density $\rho_{ij}$
has simple poles. 
In this narrow width approximation, the time evolution matrix is then
\eqqn\eIIevol{
\Uscr (t) = 
\left(
\matrix{
f_{+} (t) & {1\over\eta} f_{-} (t) \cr
\eta f_{-} (t) & f_{+} (t) \cr}\right)
}
where $f_{\pm} (t) = {1\over 2} ( V_{ss} \pm V_{ll} )=
{1\over 2} (\hbox{e}^{- {\rm i} \lambda_{S} t} \pm
   \hbox{e}^{ - {\rm i} \lambda_{L} t} ) $.
Nevertheless, since it is an approximate theory, it is not surprising 
that, for example, this procedure cannot satisfy exactly the unitarity 
requirement
\ref\kapil{B. G. Kenny and R. G. Sachs, Phys. Rev. {\bf D8} (1973) 1605;
P. K. Kabir and A. Pilaftsis, Phys. Rev. {\bf A53} (1996) 66},
which is essential for the basic interpretation of any theory. The 
assumption of constant decay rates, as they arise by the unitarity sum rules
\ref\BS{
J. S. Bell and J. Steinberger, Proceedings Oxford Int. 
Conf. on Elementary Particles 1965, ed. R. G. Moorehouse et al.
(Rutherford HEP Lab., Chilton, Didcot, Berkshire, England, 1966) p. 195} \ 
connecting the kaon system with the space of all the decaying final 
states, then, cannot be justified in an exact sense.
In fact, for instance, the modulus of the ratio of the off-diagonal elements
\eqqn\eIIxiii{
r(t) ={ {U_{12}(t)} \over {U_{21}(t)} } = \left({ p\over q}\right)^2
}
differs from unity and could vary with time~\KAK \ 
if the time-reversal $T$--invariance is not a symmetry 
of the underlying Hamiltonian. In fact, in this case,
the condition of reciprocity is not properly satisfied~\kapil .
Indeed, the inclusion of off-diagonal terms
$V_{sl}=-V_{ls}$ in the 
evolution matrix $\Vscr$, induces a modification both of the time 
evolution matrix and also of the
previous ratio. 
Assuming a global $CPT$--invariant propagation, in fact, it becomes
\eqqn\eIIxiiib{
r(t) = 
\left({ p\over q}\right)^2
\left({{1 - A}\over{1 + A}}\right) \quad.
}
On more general assumptions like causality and analycity,
the time dependence of the 
vacuum regeneration term 
$A=(V_{sl}-V_{ls})/(V_{ss}-V_{ll})$
can be obtained
to study the expansion criteria for pole dominance.
The details of the results depend, of course, upon the spectral function 
$\rho_{ij}$ defined in Eq.~\eIIxii . If the interacting sector 
$\Hscr_I$ of the Hamiltonian is small, there 
will be two (non degenerate) poles $\lambda_\alpha$ on the second 
Riemann sheet.
At this level of generalization
\ref\CDMV{D. Cocolicchio and M. Viggiano, ``{\it The Quantum Theory of 
the Kaon Oscillations}'', preprint 
IFUM FT-97, Istituto Nazionale di Fisica Nucleare,
Sezione di Milano}\ 
the time evolution matrix elements are 
given according to the following expression
\eqqn\eIIxv{
U_{ij} = \sum_{r=0}^\infty {1\over{2\pi i}} \int_{\Cscr}
e^{ -i z t} \langle K_i \vert (\left( G_0 H_I \right)^r G_0 \vert K_j 
\rangle dz
}
where $\Cscr$ is a closed curve encircling all the complex eigenvalues 
of the effective Hamiltonian $\Hscr$ and
\eqqn\eIIxvi{
\Gscr_0 (z) = { 1 \over {z \Iscr - \Hscr_0} }
}
represents the analytically continued Green's function of the non 
interacting theory.
\blankline
\leftline {\bf III. 
The Relativistic Treatment of the Neutral Kaon System.}
\blankline\noindent
In this section, we reconsider the propagator formalism of the Quantum 
Field Theory for the scalar mesons mixing, at the basis for the correct 
approach of the $\kok$ mixing.
In the absence of weak interactions, $\ko$ and $\kob$ are eigenstates 
of the strong interactions and form a degenerate 
particle--antiparticle pair in flavour state, with a common mass 
$m_\circ$ 
(whatever we assume $CPT$ invariance). When higher order weak interactions are 
introduced, transitions are induced between $\ko$ and $\kob$. Thus, 
mixing, due to quantum corrections, prohibits the $\ko$, $\kob$ scalar mesons
from propagating independently of each other. Consequently, 
the propagator of the kaon system has to be considered as a $2\times 
2$ matrix. In other words, since strangeness is not conserved in the 
kaon decays, we must consider two propagators, one for each sense of 
strangeness. Therefore, 
to describe the $\kok$ transitions induced by higher order weak interactions,
the essential tools consist in introducing four full dressed kaon 
propagators:
\eqqn\dijp{
\Delta^\prime_{ij}(k^2)\; : \qquad
\Delta^\prime_{\circ\circ}(k^2)\; , \quad
\Delta^\prime_{\circ {\overline \circ}}(k^2)\; , \quad
\Delta^\prime_{{\overline \circ} \circ}(k^2)\; , \quad
\Delta^\prime_{{\overline \circ}\ {\overline \circ}}(k^2) \quad ,
}
which can be computed perturbatively using one-particle irreducible 
self-energy parts
\eqqn\pijp{
\Pi_{ij}(k^2)\; : \qquad
\Pi_{\circ\circ}(k^2)\; , \quad
\Pi_{\circ {\overline \circ}}(k^2)\; , \quad
\Pi_{{\overline \circ} \circ}(k^2)\; , \quad
\Pi_{{\overline \circ}\ {\overline \circ}}(k^2) \quad .
}
In these expressions, the subscripts $\circ$, ${\overline \circ}$ 
correspond to incoming and outgoing eigenstates of the strong 
interactions $\ko$ and $\kob$.
The regularized propagators connecting the $\ko$ and $\kob$ states to 
themselves and to each other can be obtained according to the matrix 
Dyson equation
\eqqn\eIIIu{
\bfDelta^\prime = \bfDelta + \bfDelta\, \bfPi\, \bfDeltap
}
where the following matrix notation has been used
\blankline
\eqqn\eIIId{
\bfDelta =
\left(\matrix{\Delta & 0 \cr 0 & \Delta \cr}\right) \quad ,\quad
\bfDeltap =
\left(\matrix{ \Delta^\prime_{\circ \circ} & 
\Delta^\prime_{\circ {\overline \circ}} \cr
\Delta^\prime_{{\overline \circ} \circ} &
\Delta^\prime_{{\overline \circ} {\overline \circ}} \cr}\right)\quad ,\quad
\bfPi =
\left(\matrix{ \Pi_{\circ \circ} & 
\Pi_{\circ {\overline \circ}} \cr
\Pi_{{\overline \circ} \circ} &
\Pi_{{\overline \circ} {\overline \circ}} \cr}\right)\quad .\quad
}
If we turned off weak interactions, the bare components of the tree level 
propagator are given by
\eqqn\eIIIt{
\Delta_{ij} (k^2) =
{{\delta_{ij}} \over {k^2 - m^2_\circ}}
}
In presence of interactions, the full expression for the dressed kaon 
propagator matrix $\bfDeltap$ can be determined in principle by the 
inversion of
\eqqn\eIIIq{
\left[ \bfDeltap \right]^{-1} \simeq 
\left( \bfDelta^{-1} - \bfPi\right) =
\left(\matrix{
[k^2-m^2_\circ + \Pi_{\circ\circ} ] & \Pi_{\circ{\overline\circ}} \cr
\Pi_{{\overline\circ} \circ} & 
[k^2-m^2_{\overline\circ} + \Pi_{{\overline\circ}{\overline\circ}} ] \cr}
\right)
}
Keeping the leading terms, we find
\eqqn\eIIIc{
\eqalign{
\Delta^\prime_{\circ\circ} = &
\left[ ( k^2 - m^2_\circ + \Pi_{\circ\circ} ) -
{ { \Pi_{\circ {\overline\circ}} \Pi_{{\overline\circ} \circ} }
   \over
{ (k^2 - m^2_{\overline\circ} + \Pi_{{\overline\circ} {\overline\circ}}) } }
\right]^{-1} \quad ,\cr
\Delta^\prime_{\circ{\overline\circ}} =&
-   { \Pi_{{\overline\circ}\circ} }
\left[
{ ( k^2 - m^2_\circ + \Pi_{\circ\circ} )( k^2 - m^2_{\overline\circ} + 
\Pi_{{\overline\circ} {\overline\circ}} ) - 
\Pi_{{\overline\circ}\circ} \Pi_{\circ{\overline\circ}} }
\right]^{-1} \cr
\Delta^\prime_{{\overline\circ}{\circ}} =&
-   { \Pi_{\circ {\overline\circ} } }
\left[ { 
( k^2 - m^2_\circ + \Pi_{\circ\circ} )( k^2 - m^2_{\overline\circ} + 
\Pi_{{\overline\circ} {\overline\circ}} ) - 
\Pi_{\circ{\overline\circ}} \Pi_{{\overline\circ}\circ} }\right]^{-1} \cr
\Delta^\prime_{{\overline\circ}\ {\overline\circ}} =&
\left[ ( k^2 - m^2_{\overline\circ} + 
\Pi_{{\overline\circ} {\overline\circ}} ) -
{ {\Pi_{{\overline\circ}\circ } \Pi_{\circ{\overline\circ} } }
   \over
{ (k^2 - m^2_{\circ} + \Pi_{{\circ}{\circ}}) }}
\right]^{-1} \cr
}
}
where the dependence of the self-energies on $k^2$ is implied.
The renormalized self-energies $\Pi_{ij}(k^2)$ vanish by definition as 
$k^2\rightarrow m^2_\circ$ at least as fast as $(k^2-m^2_\circ)^2$. 
The subtleties of the higher order terms in the elements of the 
propagators are relevant only close the $(\ko\kob)_L$ resonance.
We assume that the pure fields $\ko$ and $\kob$ have already been 
renormalized in the sense that the relevant counterterms have been 
absorbed into the mass and wavefunction renormalized factors. As a 
consequence of $CPT$ invariance, the diagonal matrix elements are 
equal:
\eqqn\eIIIse{
\langle \kob \vert \bfDeltap \vert \kob \rangle =
\langle \ko \vert \bfDeltap \vert \ko \rangle 
}
whereas the off-diagonal elements $\Delta^\prime_{ij}$ are equal only in the 
case the interactions are all $CP$ invariant.
In order to describe the kaon system in terms of uncoupled channels, 
we need to diagonalize $\bfDeltap$. 
In principle, the matrix propagator
\eqqn\eIIIst{
\bfDeltap = \left[ \bfI k^2 - \bfLambda (k^2)\right]^{-1}\; ,
}
where the effective square-mass matrix is given by
\eqqn\eIIIo{
\bfLambda (k^2) = \left[ m^2_\circ \bfI + \bfPi (k^2) \right] \, ,
}
can be brought into a diagonal form
\eqqn\eIIIn{
\Delta_{\alpha\beta}^\prime (k^2) = \left( \matrix{ \Delta^\prime_S (k^2) & 0 \cr
0 & \Delta^\prime_L (k^2) \cr} \right) 
}
through a $k^2$-dependent transformation. 
We may suppose that the 
dynamics of the system is governed by 
the poles $\lambda_{S,L}$ in the propagator 
that are the complex solutions deriving from the 
vanishing of the determinant of the inverse propagator
\eqqn\SecEq{
\det (\lambda^2_\alpha \bfI - \bfLambda ) = 0 \quad .
}
This dispersion relation is just the equation which locates
the position of poles and let us write the diagonal propagator
according to the following expression
\eqqn\eIIIn{
\Delta_{\alpha\beta}^\prime
=
\left( \matrix{
{1 \over {k^2 - \lambda^2 _S}} & 0 \cr
0 & {1 \over {k^2 - \lambda^2 _L}} \cr } \right) \; .
}
Nevertheless, since in our case the coupling is weak, it is sufficient 
to approximate
\eqqn\eIIIxiii{
\Pi_{ij} (k^2) \simeq \Pi_{ij} (m^2_\circ) \; .
}
In this case,
\eqqn\eIIIxiv{
\bfLambda = \left[ m^2_\circ \bfI + \bfPi(m^2_\circ)\right]
}
represents the square of an effective complex mass matrix. It can be 
diagonalized by a similarity complex transformation $\bfR$:
\eqqn\eIIIxv{
\left[ \bfR \bfLambda \bfR^{-1} \right]_{\beta\alpha} = \lambda^2_\alpha 
\delta_{\beta\alpha}
}
where, in brief we obtain
\eqqn\eIIIxvi{
\eqalign{
\lambda^2_S = &
\frac{\displaystyle 1}{\displaystyle 2}\left[ 
\left( \Lambda_{11} + \Lambda_{22} \right) - Q \right] \cr
\lambda^2_L = &
\frac{\displaystyle 1}{\displaystyle 2}
\left[ 
\left( \Lambda_{11} + \Lambda_{22} \right) + Q \right] \cr}
}
with
\eqqn\eIIIxvii{
Q = \sqrt{
\left( \Lambda_{11} - \Lambda_{22} \right)^2 + 4 \Lambda_{12}\Lambda_{21} }
}
The complex scaling matrix $\bfR$ exists unless both $Q=0$ and the 
off--diagonal elements $\Lambda_{12}$, $\Lambda_{21}$ are different 
from zero. The physical fields $\kl$ and $\ks$ corresponding to the 
eigenvalues
\eqqn\eIIIxviii{
\lambda^2_{S, L} = \left( m_{S, L} - 
\frac{\displaystyle i}{\displaystyle 2} \gamma_{S,L}
\right)^2 \simeq m^2_{S, L} - i m_{S,L} \gamma_{S,L}
}
are combinations of the $\ko$ and $\kob$
for which only the diagonal
elements of the propagator matrix contain poles in the $\kl$, $\ks$ basis.
We define the transformation between the physical and flavour pure 
bases as usual with relations analogous to Eq. \eIIo .
Any invariance of the theory will reflect itself in an 
invariance of the propagator and then also of the square mass matrix 
$\bfLambda$. 

The fact that $\bfLambda$ is, in general,
momentum dependent does not introduce any 
additional complications, in practice, since $k^2$ is always fixed by 
the on-shell condition of the initial particles. Anyway,
the resulting eigen-physical fields are those with a definite propagation 
behaviour. 
However, disregarding, at the moment, the complications regarding the higher 
order differences among the possible schemes of renormalization, for a 
given $s=k^2$, 
we can write the regularized inverse propagator as
\eqqn\eIIIduno{
\Delta^{\prime -1}_{ij} = \left[ s \bfI - \bfLambda \right] =
\bfR_{i\alpha} \, \Delta^{\prime -1}_{\alpha\beta} \,
\bfR^{-1}_{\beta j}
}
where
\eqqn\eIIIddue{
\Delta^{\prime -1}_{\alpha\beta} =\left[ s \delta_{\alpha\beta} - 
N_{\alpha\beta} \right]
}
and
\eqqn\eIIIdtre{
N_{\alpha\beta} = {\left[\bfR^{-1} \bfLambda 
\bfR\right]}_{\alpha\beta}\quad .
}
In general, $\bfPi$ (and hence $\bfLambda$), is momentum dependent.
Expanding with respect to the $s_p$ pole of the scattering amplitude 
where $\Delta^{\prime -1}_{i j} (s_p) = 0$, we obtain
\eqqn\eIIIdqua{
\bfPi_{i j} \simeq\, \bfPi(s_p)\, +\, (s - s_p)\, \bfPi^\prime 
(s_p)\, + \, \dots 
}
Extracting the leading terms of this Laurent expansion about $s_p$,
we can write $\Delta^{\prime -1}_{\alpha\beta} $ according to
\eqqn\eIIIdcin{
\Delta^{\prime -1}_{\alpha\beta}  \simeq
{{\delta_{\alpha\beta}}  \over
{(s - s_\alpha )}}
}
where
\eqqn\eIIIdsei{
s_\alpha \simeq \lambda_\alpha^2 \left[{1 - \Pi^\prime_\alpha 
(s_p)}\right]\quad .
}
It is worth noting that $\bfLambda (k^2) $ shares all the 
properties of the effective Hamiltonian in the description of the 
kaon system. In particular, $CPT$ invariance requires that 
$\Lambda_{11} =\Lambda_{22}$ and $CP$ invariance prescribes the 
equality of the off-diagonal elements $\Lambda_{12} =\Lambda_{21}$.
Thus, the $\kok$ 
mixing gives rise to $CP$ violation through the effective mass-squared 
matrix $\bfLambda (k^2)$. The basic parameter which characterizes the 
indirect $CP$ violation induced by the mixing in the kaon system is 
given by
\eqqn\eIIIxas{
\eta = {{-q}\over p} = {{-(1-\epsilon)}\over{(1+\epsilon)}} =
\sqrt{ {\Lambda_{21}}\over{\Lambda_{12}} }
}
which is a rephasing invariant quantity and hence physically 
meaningful.

\noindent
As a further remark, we can stress that
these states of definite mass and lifetime are never 
more states with a definite $CP$ character.
The experimental evidence 
\ref\CCFT{
J. H. Christenson, J. W. Cronin, V. L. Fitch, and R. Turlay,
Phys. Rev. Lett. {\bf 13} (1964) 138}\ 
in 1964, that both the short-lived 
$K_S$ and long-lived $K_L$ states decayed to $\pi \pi$,
suggests that it is important to consider 
the $CP$ eigenstates $K_1$ and 
$K_2$, with the conventional choice of phase.
The transformation between the physical and this new basis 
can be parameterized by means of the impurity complex parameter
$\epsilon$ which encodes the indirect mixing effect of CP violation 
in the following form
\eqqn\eIIIx{
\eqalign{
\ks = &
{1 \over {\sqrt{1 + |\epsilon|^2}} }
\left( \ku - \epsilon \kd \right) 
= {1\over\sqrt{2}} \left[ (p+q) \ku - (p-q) \kd \right]
\cr
\kl = & 
{1 \over {\sqrt{1 + |\epsilon|^2}} }
\left( \kd + \epsilon \ku \right) 
= {1\over\sqrt{2}} \left[ (p-q) \ku + (p+q) \kd \right]
\cr\quad .}
}
Moreover, the propagation of this different linear 
$CP$--invariant combination of $\ko$, $\kob$ cannot be regarded as 
physical, in the sense they cannot be directly produced or detected.

\noindent
A last remark regards the fact that the most general
(non $CPT$ invariant or generally $k^2$--dependent)
transformation between the physical and the bare states 
involves two impurity $\epsilon_S\neq\epsilon_L$ factors
\eqqn\eIIIix{
\eqalign{
\ks = {1 \over {\sqrt{ 1 +|\epsilon_S|^2}} } 
& \left( \ku - \epsilon_S \kd \right) 
\cr
\kl = {1 \over {\sqrt{ 1 +|\epsilon_L|^2}} } 
& \left( \kd + \epsilon_L \ku \right)
\cr}
}
In fact, the condition that the off--diagonal components of the propagator
matrix $\bfDeltap_{\alpha\beta}$ contain no poles, will fix
$\epsilon_{S,L}$.
When the theory violates the $CPT$ invariance
and the $\Pi_{ij}(k^2)$ is $k^2$ dependent, we see explicitly 
that $\epsilon_S\neq\epsilon_L$ and the transformation from 
the bare to the physical kaons is not a simply rotation.
Furthermore, one can note that the physical masses which arise from locating
the poles in the diagonalised propagator matrix no longer
correspond to exact eigenstates.  To lowest order in the quantum corrections
the physical mass is given by $M^{\rm phys}_\circ
=\left[ m^2_\circ + \Re \Pi_{\circ\circ} ( (M^{\rm 
phys}_\circ)^2)\; \right]^{1/2}$, 
that is the real part of the pole in $\Delta^\prime_{\circ\circ}$.
Such kind of generalization of the pole mass renormalization 
scheme has been already outlined in Ref.~\Instab \ in the absence of 
particle mixing. 
In case of mass matrices, these conditions have to be fulfilled
by the corresponding eigenvalues, resulting in complicated 
expressions. These relations can be considerably simplified by 
requiring simultaneously the on--shell conditions for the 
renormalization matrices. So that, we can state that the renormalized 
one-particle irreducible two-point functions are diagonal if the 
external lines are on their mass--shell. The diagonal elements are 
then fixed such that the renormalized fields are properly normalized, 
i.e. the residues of their renormalized propagators are equal to one. 
This choice of field renormalization implies that the renormalization 
conditions for the mass parameters involve only the corresponding 
diagonal self--energies.
Assuming that $\Re \Pi_{ii} (k^2)$ vanish as $k^2\rightarrow 
m_\circ^2$, at least as $(k^2-m_\circ^2)^2$,
the following prescriptions
\eqqn\eIIIxi{
\eqalign{
\Re  \Pi_{ij} (m^2_\circ)\ = \ \Re  {\Pi}_{ji}(m^2_\circ) =& 0\, ,\cr
\lim\limits_{\displaystyle k^2\to m^2_\circ}\ \frac{1}{k^2-m^2_\circ}\, 
\Re  \Pi_{ii}(k^2) &= 0\, . \cr}
}
lead to a standard Breit-Wigner form for the propagators. Note 
that any deviation of the Breit-Wigner form and/or any non linearity 
in the $k^2$--dependence will produce a non--zero off--diagonal 
element of the propagator matrix even in the physical basis.
However, we may simply 
choose that the self-energies are regularized by requiring that the 
physical complex pole positions of the matrix elements are not shifted 
rather to impose the usual on-shell renormalization conditions.
Moreover, the inclusion of this formalism in a gauge 
theory requires the possibility to choose a pole
$s_p$ which can regularize
the self-energies to produce gauge invariant results.
In fact, the question of the correct treatment of unstable particles 
in any underlying theory faced with the problem to select gauge 
independent observables. Until the dynamics of unstable particles has 
been described in terms of initial and final asymptotic states, it 
results unitary and causal. Nevertheless, this use of on--shell 
particle configurations becomes misleading if the resummation of the 
unstable particle self-energy graphs takes into account higher--order
corrections. Nevertheless, unitarity relations, like those in Ref.~\BS ,
which no longer relate real quantities, become unclear with unstable 
states as interacting particles. Evidently, the problem of gauge 
dependence with unstable particles needs a consistent computational 
scheme which could avoid the artifacts of the resummation 
method. Actually, we neglect the non-resonant
parts of the transition amplitudes that is equivalent to consider 
isolated narrow particles. 
The resulting Breit-Wigner form of the propagator although, in 
general, is not enough to preserve gauge invariance, but it 
is expected to contain the biggest contribution of the absorptive 
part.
Namely, far from the resonant region, the decay rate of the kaon is so 
small that it is legitimate to use kaons $\ko$, $\kob$ as two 
asymptotic states of the $S$-matrix.
Within the spirit of the mass--mixing formalism we take the initial 
widths to be constant, with no explicit functional dependence on mass. 
Inclusion of such mass dependence, or working with mass rather 
mass-squared matrix, results in amplitudes changes we expected 
negligible in the limited energy range one usually works. Of course, 
this view is not justified for a very short time interval (much shorter 
than the mean life of $K_S$). In fact, in this time interval, decay 
processes cannot be described by the simple complex pole dynamics.
Finally, the question that the off diagonal elements of the self energy matrix
are momentum-dependent implies that the conventional assumption
of constant mixing ratio $r(q^2)$ remains questionable
\CDMV .
\vskip 0.5truecm
\centerline {\bf ACKNOWLEDGEMENTS}
\vskip 0.3truecm
\noindent
I wish to thank F. Botella, N. Paver and A. Pugliese for useful 
discussions and B.~Chiu, H.~Y.~Cheng and L.~Khalfin for 
correspondence.

\vfil\eject
\vfill\eject\immediate\closeout\rfile
\centerline{{\bf References}}\bigskip
\input refs.tmp\vfill\eject
 \bye